\documentclass[12pt]{iopart}
\usepackage{braket}
\usepackage{amsopn}
\usepackage{amsfonts}
\usepackage{amssymb}
\usepackage{bbold}
\usepackage{graphicx}
\usepackage{color}
\usepackage[english]{babel}
\usepackage{placeins}
\usepackage{algorithm}
\usepackage{algorithmic}
\definecolor{myblue}{rgb}{0.2,0.2,0.8}
\usepackage[colorlinks=true,citecolor=myblue,linkcolor=myblue]{hyperref}
\newcommand{\eqref}[1]{(\ref{#1})}
\DeclareMathOperator*{\argmin}{argmin}
\newcommand\newblock{\hskip .11em\@plus.33em\@minus.07em}

\begin{document}
\title{Hybrid quantum-classical optimization with cardinality constraints and applications to finance}

\author{Samuel Fern\'andez-Lorenzo}
\address{BBVA Client Solutions Research \& Patents, Calle Sauceda 28, 28050 Madrid, Spain.}
\ead{samuel.fernandez.lorenzo.contractor@bbva.com}

\author{Diego Porras}
\address{Instituto de F\'\i sica Fundamental, IFF-CSIC, Calle Serrano 113b, 28006 Madrid, Spain.}

\author{Juan Jos\'e Garc\'ia-Ripoll}
\address{Instituto de F\'\i sica Fundamental, IFF-CSIC, Calle Serrano 113b, 28006 Madrid, Spain.}

\date{\today}

\begin{abstract}
In this work we develop tools to address combinatorial optimization problems with a cardinality constraint, in which only a subset of variables end up having nonzero values. 
Firstly, we introduce a new heuristic pruning method that iteratively discards variables through a hybrid quantum-classical optimization step. Secondly, we analyse the use of soft constraints in the form of ``chemical potentials" to control the number of non-zero variables. 
We illustrate the power of both techniques using the problem of index tracking, which aims to mimicking the performance of a financial index with a balanced subset of assets. 
We also compare the performance of different state-of-the-art quantum variational optimization algorithms in our pruning method. 
\end{abstract}


\maketitle
\section{Introduction}
Many relevant problems in quantitative finance translate into daunting computational tasks, such as combinatorial optimization problems and Monte Carlo simulations\ \cite{Brandimarte2013}, suffering from lack of parallelization or slow convergence. The emergent field of quantum finance, rooted in quantum physics and quantum computing, develops new algorithms and formulations of financial problems, to address these problems and limitations\ \cite{Orus2019}. Recent works in quantum finance have explored the design of optimal trading trajectories\ \cite{Rosenberg2016a}, credit scoring\ \cite{Milne2017}, portfolio optimization\ \cite{Rebentrost2018,kerenidis2019,hodson2019,barkoutsos2020}, Monte Carlo pricing of derivatives\ \cite{Rebentrost2018a}, risk analysis\ \cite{Woerner2019} or financial crisis forecast\ \cite{Orus2019a}, among others.

Among these challenges, in this work we focus on the family of quadratic optimization with cardinality and linear constraints, which we write as
\begin{eqnarray}
&\argmin_{\mathbf{w}} T_\mathrm{err} :=\argmin_{\mathbf{w}} \left[\mathbf{w}^T\Sigma\mathbf{w}-2\mathbf{w}^T\mathbf{g}+\epsilon_0\right] ,\label{original_problem} \\
&||\mathbf{w}||_0=\sum_{j=1}^N |w_j|^0= d, \nonumber\\
&A\mathbf{w}\leq \mathbf b \nonumber. 
\end{eqnarray}
Here, the $w_j\in(-\infty,\infty)$ are $N$ weights of which only $d<N$ are nonzero, $\Vert{w}\Vert_0=d,$ using 0-norm notation, and $\Sigma$ being a positive semi-definite matrix. Equality and inequality linear constraints are condensed in the term $A\mathbf{w}\leq \mathbf b$. 

Many canonical financial problems can be cast into the form \eqref{original_problem}, like portfolio selection in a mean-variance framework \cite{Moral-Escudero} or  \textit{index tracking} \cite{ruiz2009}, where a benchmark index with $N$ possible securities is approximated by a basket of $d\ll N$ assets properly selected and weighted. Beyond finance, we may encounter problem \eqref{original_problem} in other disciplines such as machine learning---like in future selection of a multi-variate linear regression---or supply-chain optimization. In this work, we shall rely on a simplified version of the index tracking problem as a workbench.
Problem \eqref{original_problem} poses an implicit mixed-integer optimization as the selection of $d$ variables is already a combinatorial optimization whose state space grows very fast as ${N\choose d}$. 
The limit $N, d \gg 1$, and $N \gg d$, is specially relevant for many practical optimization tasks, including the index tracking problem. 
In this limit, the number of possible solutions actually grows exponentially as a function of the number of total variables, showing a dependence $\sim N^d$. Exploring all configurations by brute force quickly becomes unfeasible, like in the case of financial problems composed by hundreds or thousands of securities, thereby approximation methods \cite{zheng2014,monge2017} and heuristic approaches come into play\ \cite{jansen2002,beasley2003,coleman2006,ruiz2009,gilli2002,varsei2013}. In particular, when addressing this problem in Noisy Intermediate Scale Quantum (NISQ) devices\ \cite{preskill2018}, a typical route is to discretize the weights and transform it into a Quadratic Unconstrained Binary Optimization (QUBO) problem with constraints. Unfortunately, this leads to an algebraic growth in the number of qubits which severely limits the size and interest of real instances that can be addressed.

Our solution to problem\ \eqref{original_problem} is a hybrid quantum-classical algorithm that we call \textit{heuristic k-step pruning} (k-PA). This method nests two interdependent optimization problems: a classical convex optimization stage to determine the relative weights of the assets, combined with a quadratic binary optimization that determines the relevant subset of $d$ variables that optimizes \eqref{original_problem}. This approach makes best use of small NISQ devices, maximizing the number of weights that are described by a single quantum register---namely, $N$ can include a universe of assets with up to $N$ securities. We demonstrate that the pruning method can be executed in NISQ quantum computers (e.g. IBM-Q), using variational methods with a very small number of gates but very good performance. However, the same method can be extended to work with quantum annealers (D-Wave's) or quantum-inspired optimizers. 

We stress that our pruning algorithm uses a classical computer for the convex optimization step, 
which is not computationally hard, whereas it uses a quantum computer to address the hard combinatorial problem of selecting the $d$ variables that take non-zero values. 
Thus, our algorithm ``concentrates'' the hardness of combinatorial optimization into the step that is solved by the quantum computer. 
Our algorithm has a further advantage since, even though the original problem is stated in terms of continuous variables, the quantum computer is used precisely in the step where that problem is reduced to binary variables, which can be more efficiently stored and processed by qubits. 
In principle, the variable selection step could also be solved by means of classical QUBO solvers. However, in applications beyond hundred or thousand variables, it is computationally challenging to find the global optimum, and here quantum computers may offer an advantage over classical methods.
Since we expect that near or mid-term quantum devices will have severe limitations in size, the advantages offered by our scheme can be decisive for finding useful practical applications of those devices. 

In Section \ref{iterative_pruning} we describe and analyze our first contribution, the pruning algorithm. In Section \ref{constraints_section} we discuss the introduction of constraints in optimization problems. We compare the use of hard constraints with an approach based on Lagrange multipliers or ``chemical potentials''. Section \ref{benchmarks} introduces the variational ans\"{a}tze and classical optimizers employed in our study, as well as the index tracking problem that we will use to benchmark our solutions. Our algorithms and existing techniques are then numerically compared in Section \ref{results} by using a real dataset of intra-day data of the PHLX Oil Service Sector (OSX) index, composed of 15 companies involved in the oil services sector (cf.\ \ref{fig:index}). 
Finally, Section  \ref{conclusions} discusses the conclusions of our study.

%
\section{Iterative elimination of variables} \label{iterative_pruning}

\subsection{One-step pruning}

The optimization problem from Eq.\ \eqref{original_problem} is a mixed-integer optimization that implicitly joins a computationally hard task---the selection of $d$ variables within an exponentially large family of choices---, with a classically tractable, convex optimization problem---assigning the weights of the $d$ selected variables $\{w_{j_n}\}_{n=1}^d$---. This suggest that we split the whole problem into two problems, which we later address by a quantum and a classical computer, in line with the current philosophy of hybrid classical-quantum computations.

The first iteration of our idea is the \textit{single-step selection algorithm} (1-SA). Let us define a vector $\mathbf{x}$ of decision variables, which are nonzero $x_i=1$ only when the corresponding weight $w_i$ is included in the final basket of variables. We write down the combinatorial optimization problem of selecting the $d$ variables that will have non-zero weights
\begin{eqnarray} \label{selection_problem}
&\argmin_{\mathbf{x}} S_\mathrm{err} := \argmin_{\mathbf{x}} \left[\mathbf{x}^T\Sigma\mathbf{x}-2\mathbf{x}^T\mathbf{g}\right] \\
&||\mathbf{x}||_0= d. \nonumber
\end{eqnarray}
The solution of this is used to build a cost function with smaller matrices, $\Sigma_r(\mathbf{x})$ and $g_r(\mathbf{x}),$ that only contain the rows and columns where $x_i\neq 0.$ The reduced problem gives us a smaller vector of weights $\mathbf{w}_r\in \mathbb{R}^d$ by solving the following convex optimization problem
\begin{eqnarray}  \label{reduced_convex}
&\argmin_{\mathbf{w}_r} T_\mathrm{err}^{r}(\mathbf{x}) :=\argmin_{\mathbf{w}_r} \left[\mathbf{w}_r^T\Sigma_{r}(\mathbf{x})\mathbf{w}_r-2\mathbf{w}_r^T\mathbf{g}_{r}(\mathbf{x})+\epsilon_0 \right] \\
&A\mathbf{w}\leq \mathbf{b}. \nonumber
\end{eqnarray}

This 1-SA formulation is problematic because the selection of variables does not consider their relative importance in the final weights. A better strategy is the \textit{single-step pruning algorithm} (1-PA). In this reversed strategy, the cost function of the selection problem incorporates the weights, through the transformation $x_i\rightarrow w_ix_i.$ More precisely, we first solve the convex optimization problem
\begin{eqnarray} \label{full_convex}
&\argmin_{\mathbf{w}} T_\mathrm{err}^{full} :=\argmin_{\mathbf{w}} \left[\mathbf{w}^T\Sigma\mathbf{w}-2\mathbf{w}^T\mathbf{g}+\epsilon_0\right]  \\ \label{convex_problem_full}
&A\mathbf{w}\leq \mathbf b. \nonumber
\end{eqnarray}
After solving Eq. \eqref{full_convex}, we obtain a vector of optimal weights, ${\bf w}_{opt}$. This information is subsequently incorporated into a diagonal matrix $D_{ij}({\bf w}_{opt})= 
w_{opt,i} \delta_{ij},$ with which we select the variables in a modified purely combinatorial optimization problem
\begin{eqnarray}
&\argmin_{\mathbf{x}} P_\mathrm{err}(\mathbf{w}_{opt}) :=\argmin_{\mathbf{x}} \left[\mathbf{x}^T\mathbf{D}(\mathbf{w}_{opt})\Sigma\mathbf{ D}(\mathbf{w}_{opt})\mathbf{x}
-2\mathbf{x}^T\mathbf{D}(\mathbf{w}_{opt})\mathbf{g}\right] \label{singlepruning}\\
&||\mathbf{x}||_0= d. \nonumber
\end{eqnarray}
Finally, we update the optimal weights for this combination through another convex optimization problem on a reduced subspace like in Eq.\ \eqref{reduced_convex}.

\subsection{k-step pruning}

We can use 1-PA as the basis for an \textit{iterative pruning algorithm} that constructs solutions for decreasing basket sizes. More precisely one would design a discrete schedule with $k$ steps and decreasing universe sizes\footnote{One possibility is to decrease the size of the search spaces linearly, $N_{i}=N-s\times i,$ but other schedules are also feasible.} $N_1=N > d_1=N_2 > \ldots \geq N_{k} > d_k=d.$ For the $k$-th step, the 1-PA algorithm assigns a series of weights to all variables in the universe of size $N_k,$ constructs the reduced matrices $\Sigma_{r}$ and $g_{r}$ and selects a basket of $d_k$ nonzero variables that will move on to the next step.

We can combine the iterative pruning with stochastic optimization methods, such as a quantum optimization algorithm, to solve each of the 1-PA variable selection steps. In this scenario, we have to fine tune not only the universe size, but also the number of repetitions of the stochastic method. We suggest to gradually increase the number of repetitions as the universe size shrinks, according to the schedule $r\rightarrow r_0 +\alpha r.$ Heuristically, in early stages where $d\sim N,$ we are more likely to retain variables in the optimal portfolio, so it makes sense to spend less repetitions. In later stages, each repetition will be exponentially cheaper and more accurate, due to the decrease in universe size, and it pays off to use the same number of resources to get more accurate intermediate solutions. We refer to the combination of universe size and repetition schedule as our heuristic k-step pruning algorithm (k-PA). It is described in Algorithm\ \ref{algorithm_def} as a pseudocode function to calculate a target basket of size $d_\mathrm{target}$ and weights $w_j$ from an universe of size $N$ in steps of size $s.$

\begin{algorithm}[h!]
\renewcommand{\algorithmicrequire}{\textbf{Input:}}
\renewcommand{\algorithmicprint}{\textbf{break}}
\caption{heuristic k-step pruning algorithm}
\begin{algorithmic}\label{algorithm_def}
\REQUIRE $N,d_\mathrm{target},r_0,\alpha,s$
\STATE $d \leftarrow N$
\STATE get ideal weights $w_j=w_{opt,j}$ or find them from ${\argmin_{\mathbf{w}}} T_\mathrm{err}^{full}$
\WHILE{True}
	\STATE $r \leftarrow r_0+\alpha r$
	\STATE $d \leftarrow \max(p,d-s)$
	\FOR{1:r}
		\STATE \textit{approximately} solve ${\argmin_{\mathbf{x}}}  P_\mathrm{err}$ with weights $w_j,$ to select $d$ variables
		\STATE keep the best solution $\mathbf{x}$
	\ENDFOR
	\STATE solve ${\argmin_{\mathbf{w}}}T_\mathrm{err}^{r}(\mathbf{x})$  with linear constraints for selected variables in $\mathbf{x}$ to obtain new optimal weights $w_{opt,j}$
	\STATE construct $\Sigma_{r}(\mathbf{x})$ and $g_{r}(\mathbf{x})$
	\STATE $N \leftarrow d$
	\IF{$N=d_\mathrm{target}$}
		\PRINT
	\ENDIF
\ENDWHILE
\end{algorithmic}
\end{algorithm}

In Algorithm\ \ref{algorithm_def}, $r_0$, $\alpha$ and $s$ are hyperparameters that have to be chosen depending on the specific problem at hand. 
Parameter $s$ is particularly relevant, since it determines the number of pruning steps, as well as the dimension of the configuration space at each step. 
The value of $s$ should be large enough so that at each step of the pruning process,
a subset of weights are selected that takes into account significant correlations with the rest of the weights. 
However, a large value of $s$ can also induce errors related to the reduction of the original weights, $w_j$, to binary values, $x_j$. 
The optimal value of $s$ will thus inevitably have to be calculated for particular problems by directly applying our method and comparing the quality of the solutions.

\section{Optimization with constraints}\label{constraints_section}

The pruning algorithm consists of two stages: a classically simple step in which the $N$ variables are assigned prospective weights by convex optimization, and a second one involving a quadratic binary optimization to select $d$ nonzero variables\ \eqref{singlepruning}. We will now discuss two methods to treat this second stage.

\subsection{Hard-constraint formulation} \label{hard_constraint}

Quantum variational algorithms where originally devised for tackling unconstrained optimization problems. One alternative to circumvent this limitation is to include the cardinality constraints in Eq. \eqref{singlepruning} into a Quadratic Unconstrained Binary Optimization (QUBO) cost function, as quadratic penalties.
\begin{equation} \label{QUBO}
\argmin_{\mathbf{x}} P_\mathrm{err}^{QUBO} = \argmin_{\mathbf{x}}  \mathbf{x}^T\mathbf{D}\Sigma\mathbf{ D}\mathbf{x}-2\mathbf{x}^T\mathbf{Dg}+P\left(\mathbf{1}^T\mathbf{x} - d\right)^2.
\end{equation}
Here, $\mathbf{1}$ is a vector of ones and $P$ is a penalty weight---large enough to prevent unfeasible configurations, but small enough to allow tunneling between different baskets that satisfy the constraints. The optimization problem\ \eqref{QUBO} can be brought to a quadratic form $P_\mathrm{err}^{QUBO} \sim \mathbf{x}^T Q \mathbf{x},$ including the linear terms\footnote{Note that for binary variables $x_i^2 = x_i x_i = x_i$ and
  $\sim \mathbf{b}^t \mathbf{x}=\sum_{ij}x_ix_j b_i\delta_{ij}.$}. The QUBO formulation with hard constraints allows us to control very well the basket size, but it creates very rough energy landscapes where classical optimizers such as COBYLA get easily trapped. For that reason, we introduce now a technique inspired in the notion of ``chemical potential'' from statistical and condensed matter physics.

\subsection{Soft-constraint formulation} \label{soft_constraints}

We can improve the convergence of the optimizer by relaxing our control over the basket size, changing the cost function to include Lagrange multipliers instead of quadratic penalties. The new regularization term acts as a \textit{chemical potential} that favors different values of the constraint, depending on the regularization parameter $\lambda$,
\begin{equation}
P_\mathrm{err}^{\mathrm{QUBO}_\mathrm{soft}} \sim \mathbf{x}^T Q \mathbf{x}+\lambda||\mathbf{x}||_0.
\end{equation}
Although regularization is widely used in machine learning (such as lasso and ridge regularization), this particular form is not so common, because it leads to non-differentiable functions and NP-hard formulations. In our case the regularization amounts to introducing a multiple of the identity matrix $\mathbb{1}$
\begin{equation} \label{QUBOsoft}
\argmin_{\mathbf{x}} P_\mathrm{err}^{\mathrm{QUBO}_\mathrm{soft}} =\argmin_{\mathbf{x}}\left[\mathbf{x}^T\mathbf{D}(\Sigma+\lambda\mathbb{1})\mathbf{ D}\mathbf{x}-2\mathbf{x}^T\mathbf{Dg}\right],
\end{equation}
updating the diagonal elements of $Q.$ Note that in the case of financial applications, $\lambda$ can also be interpreted as a constant transaction cost for each security. The cardinality constraint is thus not explicitly imposed on Eq. 
\eqref{QUBOsoft}, but the term $\lambda$ is used to control the number of non-zero variables.

A drawback of this method is that we cannot predict \textit{a priori} the number of variables $d(\lambda)$ selected by the regularization parameter $\lambda.$ In practical applications, such as index tracking, this is not a problem, because the same optimization can be repeated multiple times, with minor variations in the data, but not the structure of the problem. Under such circumstances, one can find a monotonic relationship between the regularization parameter and the average number of variables selected (as exemplified in figure \ref{fig9}).

\section{Benchmarks} \label{benchmarks}

In the remaining of this work we will study the application of the pruning, with hard and soft constraints, to a model problem of index tracking that uses data from a real-world index. For the variable selection stages we will explore the use of both quantum and classical algorithms. In the following sections we describe the quantum variational algorithm, with the ans\"atze that we will apply, the classical algorithms that we compare with, and the specific instance of the index tracking problem which we use to define our numerical simulations.

Applications of quantum algorithms to similar problems, like portfolio optimization have been published in the last years. 
Quantum algorithms which offer a speed up in portfolio optimization can be found in \cite{Rebentrost2018, kerenidis2019}, although they are not specifically designed for the index tracking problem discussed here. Other algorithms for portfolio optimization using hybrid quantum-classical methods like \cite{hodson2019, Woerner2019} are closer to our own approach. The distinctive feature of our approach is that our scheme requires fewer qubits and thus it is more feasible with near term quantum devices.

\subsection{Index-tracking problem}
\label{index_tracking_problem}
\begin{figure}[t]
  \centering
  \includegraphics[width=0.65\textwidth]{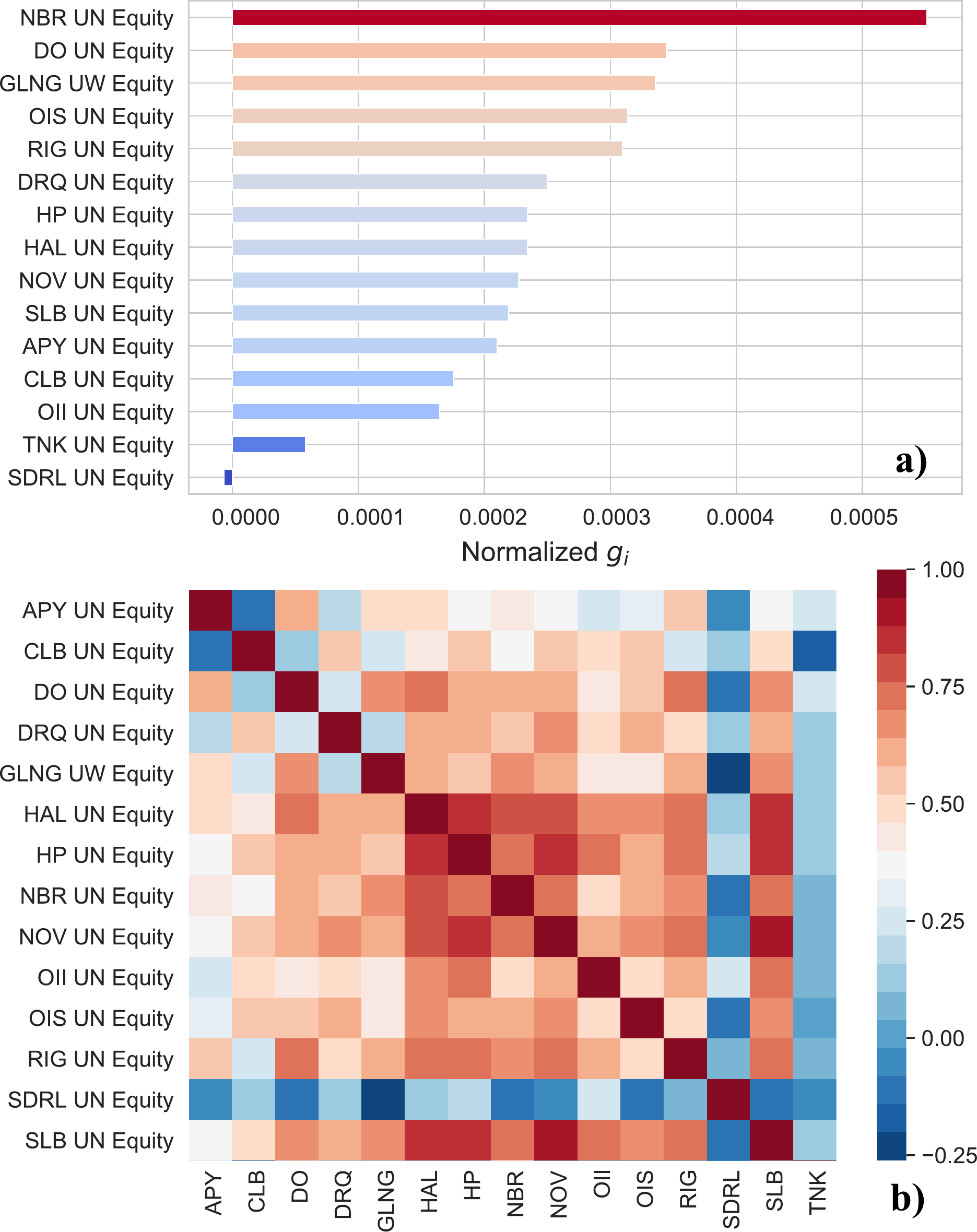}
  \caption{Composition of the PHLX Oil Service Sector (OSX) index, with look-back window on 2019-06-17. (a) Names of the assets and weights on the normalized linear components $g_i.$ (b) Correlation heatmap between equities, related to the quadratic form $\Sigma_{ij}.$}
  \label{fig:index}
\end{figure}

The goal of the index tracking problem is to replicate, over a given look-back window, a benchmark index by selecting a portfolio of $d$ components out of a large universe $N\gg d$ of possible assets. We expect that the smaller tracking portfolio will closely match the returns of the benchmark index, while at the same time reducing the transaction costs and simplifying the management process required to keep both the portfolio in-sync with the index (yet, excessive low values of $d$ would result in poor tracking). We will now show that this goal can be mapped to a particular case of the generic optimization task \eqref{original_problem}.

Let us assume that we know the prices $P_j(t)$ of the assets in our universe over an interval of time $t\in[0,T],$ which we divide into periods with fixed size $t_n=n\times \Delta{t}.$ At any time, the composition of our portfolio is determined by a set $c_j(t)$ of product units that are adapted during the tracking period. The value of the portfolio during this interval is
\begin{equation}
P(t) = \sum_{j=1}^d c_j P_j(t), \quad t \in [0 , T].
\end{equation}
Within each sub-interval, we can define the overall return of the portfolio $r_p(t_n)$ as the relative increment in portfolio value. It can be related to the returns of individual assets $r_i(t_n)=P_i(t_n+\Delta{t})/P(t_n)-1$
\begin{equation} \label{portfolioReturn}
r_p(t_n)=\frac{P(t_n + \Delta t)-P(t_n)}{P(t_n)} = \sum_j^N w_j  r_j(t_n) = \mathbf{w}^T\mathbf{r}(t_n),
\end{equation}
through a vector of weights $\mathbf{w}\in\mathbb{R}^N$ characterizing the portfolio
\begin{equation}
w_j=\frac{c_j(t) P_j(t)}{\sum_j^n c_j(t) P_j(t)}\in[0,1].
\end{equation}

The benchmark index $I(t)$ is known through a time series of returns, $r_I(t_n)=I(t_n+\Delta{t})/I(t_n)-1,$ taken by consecutive snapshots at periodic intervals. We will use this information, to construct the portfolio with $d$ assets (and fixed weights) that best replicates the index' trajectory\ \cite{ruiz2009}. A reasonable assumption is that this optimal portfolio will perform similarly to the index over a short time horizon, $t\in [T, T + T_H]$. A more sophisticated approach would involve a forecast of future returns of each index component, and to optimize our tracking portfolio over such forecast instead of optimizing over historical returns. 

To implement the search of the optimal portfolio, we introduce a tracking error $T_\mathrm{err}$ that measures the \textit{distance} between the sequence of returns of the index and our portfolio. We use the mean squared error between both time series, a quadratic form
\begin{equation}
\label{mixed_integer}
T_\mathrm{err}=\sum_{n}^{T}(r_p(t_n)-r_I(t_n))^2 =
\sum_{i,j = 1}^N w_i \Sigma_{i j} w_j - 2 \sum_{j=1}^N w_j g_j + \epsilon_0,
\end{equation}
with a matrix $\Sigma_{ij},$ a vector $g_i$ and an offset $\epsilon_0,$ defined as
\begin{equation}
\Sigma_{i j} = \sum_{n=0}^{n_T} r_i(t_n)r_j(t_n),\quad g_j = \sum_{n=0}^{n_T} r_j(t_n) r_I(t_n),\quad\epsilon_0 = \sum_{n=0}^{n_T} r_I(t_n)^2.
\end{equation}
The index can also be expressed in terms of a vector of weights $\mathbf{W} \in\mathbb{R}^{N}$ valid for the specified time interval\footnote{Even if those weights are not explicitly provided or available in financial data, such vector can be computed by convex optimization of the universe of $N$ index assets with no other restriction that demanding positive weights for all of them and proper normalization (budget constraint).}, giving us explicit formulas for the components as
\begin{equation}
g_i = \sum_{j=1}^N \Sigma_{ij}W_j\quad\epsilon_0 = \sum_{i,j=1}^N W_i\Sigma_{ij}W_j.
\end{equation}
The search for the optimal portfolio results in the following mixed-integer optimization problem 
\begin{eqnarray}
&\argmin_{\mathbf{w}} T_\mathrm{err} :=\argmin_{\mathbf{w}} \left[\mathbf{w}^T\Sigma\mathbf{w}-2\mathbf{w}^T\mathbf{g}+\epsilon_0\right] \label{cost_function} \\
&||\mathbf{w}||_0=\sum_{j=1}^N |w_j|^0= d,  \label{cardinality} \\
&||\mathbf{w}||_1=\sum_{j=1}^N |w_j|= 1, \\
&w_j\geq 0 \quad\forall j .  \label{short}
\end{eqnarray}
Here, the $w_j\in[0,1]$ are $N$ normalized weights with $\Vert{w}\Vert_1=1,$ of which only $d$ are nonzero, $\Vert{w}\Vert_0=d,$ using the 1-norm and 0-norm notations. Eq.\ \eqref{short}  is introduced to avoid short selling, and the integer nature of the problem is manifested when selecting a portfolio of $d$ assets so that Eq.\ \eqref{cardinality} is respected. Notice that the index tracking problem here defined is an instance of the generic optimization problem \eqref{original_problem}.

Our formulation assumes that $\mathbf{w}$ is a vector of continuous values, which is a good approximation commonly employed in practice. However, one could also tackle the problem as fully combinatorial recalling that equities are bought and sold in fixed lot sizes, thereby requiring a discretization in the values of $w_j$ as discussed in\ \cite{WareCorp}.

Notice that the symmetric matrix $\Sigma_{ij}$ is similar to a covariance matrix among assets,
whereas the vector $g_j$ accounts for covariance-like components with respect to the benchmark.
Hence, the cost function\ \eqref{cost_function} can be interpreted similarly to a Markowitz's portfolio optimization: the mission of the quadratic term is to reduce the volatility associated to the uncertain evolution of the assets while the linear term favors those assets more correlated with the benchmark\ \cite{moral2006,Ruiz-Torrubiano2010}.

The problem, as stated, tends to favor assets with larger weights in the index. In realistic scenarios there may be further constraints to ensure a well-diversified portfolio, such as a capital concentration constraint of the form,
\begin{equation} \label{concentration_constraint}
\mathbf{l}\leq C\mathbf{w}\leq \mathbf{u},
\end{equation}
where $\mathbf{l}$ and $\mathbf{u}$ are vectors indicating the lower and upper bounds of certain linear combinations of investments gathered in the matrix $C$. Although this type of constraint is also linear and can be tackled by our heuristic pruning algorithm, we shall not consider them in this work. Instead, we focus on drawing the quantities $\Sigma_{ij}$ and $g_{i}$ from a data set that leads to a computationally demanding problem. More precisely, we will look for indices with assets that are strongly correlated, in a way that makes the quadratic terms are comparable to the linear terms. Otherwise, the problem can be solved in good approximation retaining just the linear contribution, which comes down to sorting assets by their contribution to the index and picking the top $d$ from the list. For our numerical studies, we picked an index consisting of a cluster of 15 oil companies: the PHLX Oil Service Sector (OSX) index ($N=15$ consequently). The dataset gathers prices of all the instruments with interval $\Delta t$ of 20 minutes over several days. We take look-back windows corresponding to a single day, and average over several days. An example of the correlation heat map of these equities and their corresponding (normalized) component $g_{i}$ for a random day (2019-06-17) is shown in figure\ \ref{fig:index}.

Let's clarify that such a small index tracking problem can be efficiently solved even by brute force. We use this problem simply as a workbench that allows us to classically simulate small quantum variational circuits and to demonstrate our algorithms. In a real scenario, a financial index composed of hundreds or thousands of securities along with capital concentration constraints would pose a real computationally hard task. Since we expect NISQ devices with hundreds or thousands over the next years, our work is intended to illustrate potential use cases of this technology. 

\subsection{Quantum variational algorithms}

We shall now address how to solve the integer optimization problem from Eq.\ \eqref{singlepruning} using variational quantum circuits for combinatorial optimization\ \cite{Farhi2014,Farhi2016, McClean2016,Moll2018,Zhou2018,Rieffel2019}. Generally speaking, these methods represent the solution of the optimization problem as a quantum superposition over all possible configurations. The wavefunction is parameterized by the angles and phases of the quantum gates used to build it. These real parameters are processed by some classical optimizer, in a hybrid classical-quantum model that starts from rather uniform superpositions and aims at concentrating probability on the best arguments to our cost function. The use of quantum variational methods opens new questions, such as the shape and structure of the ansatz, the type of classical optimizers to use, or the role of quantum fluctuations and entanglement in exploring the complex and noisy space of solutions.

We can write the QUBO cost function as an Ising model,
\begin{equation}
H_\mathrm{err} = \sum_{ij} J_{ij}\sigma^z_i\sigma^z_j + \sum_i \sigma^z_i h_i + E_0,
\label{Ising}
\end{equation}
by means of the mapping $x_i \rightarrow \frac{1}{2}(1-\sigma^z_i),$ from bits to Pauli Z operator $\sigma^z_i.$ Our optimization problem translates to finding one of the ground states $\ket{s_1,s_2,\ldots,s_N}$ of $H_\mathrm{err}.$ We will approximate this as a variational method
\begin{equation}
\label{eq:variational}
\argmin_{\vec{\theta}} \braket{\Psi(\vec{\theta})|H_\mathrm{err}|\Psi(\vec{\theta})}
\end{equation}
finding the best choice $\vec{\theta}$ over a family of wavefunctions $\ket{\Psi(\theta)},$ constructed with gates that are parameterized by the angles $\theta_i.$ The optimization is done classically, constructing the wavefunction in the quantum computer, estimating the energy and estimating an update for the parameters. 

We expect that, if the variational family is dense enough, $|\Psi (\vec{\theta}\rangle)$ 
will approximate the ground state with arbitrary accuracy, or at least the spin configuration that 
minimizes $H_{\rm err}$ can be detected with a
significant probability after sampling  $|\Psi (\vec{\theta}\rangle)$.
In all the numerical experiments below we assume that variational wavefunctions are sampled by repeating experiments a number of times $N_{\rm meas} = 100$.

\subsection{Variational states}\label{variational_states}

Given the limitations of NISQ devices concerning circuit depth, a first approach is to restrict ourselves to hardware-efficient trial states\ \cite{McClean2016}, variational wavefunctions created within the limitations of existing devices. More precisely, we will use the VQE Ry ansatz (henceforth just VQE) introduced in Ref.\ \cite{Kandala2017}. Formally, the wavefunction reads
\begin{equation}
 |\Psi_{VQE} (\vec{\theta}) \rangle  =
 \left(\prod_{j=2}^p  \exp\left(-i\sum_l \theta_{jl}\sigma^y_i\right) U_{ent} \right) \exp\left(-i\sum_l \theta_{1l}\sigma^y_i\right) \ket{\Psi_0}.
\end{equation}
The ansatz starts from the zero state of the quantum register
$\ket{\Psi_0}= \ket{0}_1 \cdots \ket{0}_N,$ and then subsequent layers of local rotations and entangling operations are applied. The variational parameters $\theta_{jl}$ determine the local rotation on the $l$-th qubit in the $j$-th layer. As entangling operations we use a sequence of control-Z gates between nearest-neighbor qubits, assuming the qubits form a 1D register.

The Quantum Approximate Optimization Algorithm\ \cite{Farhi2014} (QAOA) is more hardware-demanding ansatz, that implements Trotterized version of quantum annealing. The variational parameters are the angles $\beta_n$ of global x-rotations and of $\gamma_n$ entangling operations implemented by the problem Hamiltonian,
\begin{equation}
\ket{\Psi_{QAOA} (\vec{\gamma},\vec{\beta})}  =
\prod_{n=1}^p \exp\left(- i \beta_n \sum_i \sigma^x_i \right) \exp \left(- i \gamma_n H_\mathrm{err} \right)
\ket{\Psi_+}.
\end{equation}
The initial state is the uniform superposition of all register states $| \Psi_+ \rangle = \bigotimes_{j=1}^N |+\rangle_j,$ with $|+\rangle_j = (1/\sqrt{2}) (|0 \rangle_j + |1\rangle_j ).$ In the limit of very large number of layers, QAOA may converge to the dynamics of a quantum annealing process and therefore prepare the ground state with high fidelity. In practice, the number of layers $p$ is smaller, but the ansatz captures the structure of the problem, which is why it is expected to perform better than the generic VQE method, but with an added cost of more gates to implement $H_\mathrm{err}$.

Another strategy is to force the quantum circuit to search over subspaces of the Hilbert spaces that respect the cardinality constraint  without the need of imposing an energy penalty.
The idea is to start with an initial wavefunction with a well defined number of nonzero qubits and generate the variational ansatz with a sequence of quantum gates which do conserve spin-excitation number\ \cite{Hadfield2019}.
\begin{eqnarray}
&&\ket{\Psi_{\mathrm{SWAP},d}(\vec{\theta})}  =
\nonumber \\
&&\prod_{j=2}^p \left[\exp\left(-i \sum_i \theta_{ji} \sigma^z_i \right) U_{\sqrt{\mathrm{SWAP}}} \right] \exp\left(-i \sum_i \theta_{1i} \sigma^z_i \right)
\ket{\Psi(d)}.
\end{eqnarray}
The initial state $\ket{\Psi(d)}$ has $d$ qubits with value $1,$ equispaced over the quantum register. Local rotations around the $z$-axis are interleaved with $U_{\sqrt{\mathrm{SWAP}}}$ gates that implements a 50-50 beam-splitter, partially swapping the qubit excitations
\begin{equation}
U_{\sqrt{\mathrm{SWAP}}} = \prod_{j=1}^N \exp \left( i \frac{\pi}{4} \left( \sigma^+_j \sigma_{j+1}^- + \sigma^+_{j+1} \sigma_{j}^-  \right)  \right).
\end{equation}

\subsection{Classical Optimization}

As for the classical optimization part, derivative-free optimizers like Constrained Optimization BY Linear Approximation (COBYLA) have proven to be more effective and robust in this task than gradient-based optimizers like gradient descent\ \cite{nannicini2019}. Nevertheless, all these methods tend to easily get stuck in local minima when the parameter landscape is highly nonconvex. 
We checked that this problem is specially acute for the case of low-depth QAOA, in which all the energy structure plus the energy barriers penalizing unfeasible combinations get "compressed" into a few parameters. 
An example of the poor performance of COBYLA when working with QAOA can be found in our Fig. \ref{fig7}a below
For that reason, we chose the implementation of a global derivative-free optimizer based on dual annealing from scipy in Python, which combines a generalization of simulated annealing coupled to a local search. Unlike COBYLA, this method finalizes when the maximum number of iterations is reached; this is limited by the parameter $maxiter,$ which regulates the maximum number of global search iterations.

When using COBYLA, the parameter $tol$ marks a convergence criteria; we found that $tol=0.01$ was enough to ensure convergence and avoid unnecessary iterations at the same time. Yet, the method has to be limited by a total number of iterations, which was set to $2000$ because our numerical experiments normally converge within that limit. To attempt a fair comparison with dual annealing, we should specify a $maxiter$ parameter provided by this method from which the quality of results plateaus. Finally, we have to establish some bounds where the global optimizer is going to be searching: we set the intervals $[0,2\pi]$ and $[0,\pi]$ for $\vec{\gamma}$ and $\vec{\beta}$ respectively. No further fine tuning was considered apart from the already mentioned.

\section{Numerical results} \label{results}

\subsection{Utility of heuristic pruning algorithm}

\begin{figure}[ht]
  \centering
  \includegraphics[width=0.65\textwidth]{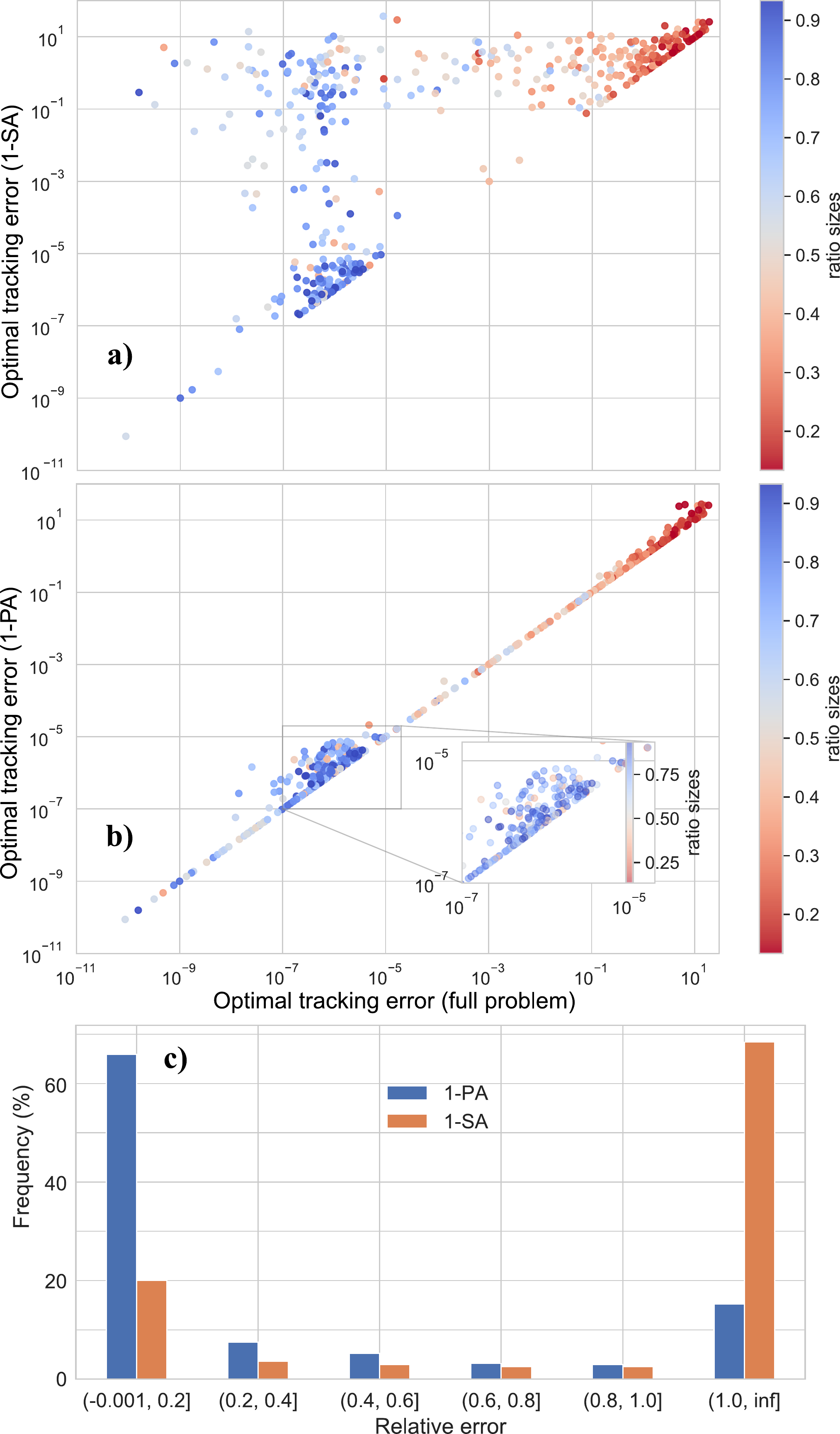}
  \caption{
  Single-step selection and pruning algorithms. 
  Global minimum tracking error using: 
  (a) 1-SA and (b) 1-PA vs. the global optimal tracking error of the problem obtained by an exact, brute force, optimization method. Calculations have been carried out for random days and different ratios $d/N$, with $N = 15$ assets taken from the PHLX Oil Service Sector (OSX) index (see Subsection \ref{index_tracking_problem} for details). 
  (c) Numerical probabilities of achieving a certain relative tracking error $\Delta=(T_\mathrm{err}-T_\mathrm{err}^\mathrm{opt})/T_\mathrm{err}^\mathrm{opt}$ in 1-SA and 1-PA, when compared with the exact optimal value, calculated by an exact, brute force, optimization.
  }
  \label{fig2}
\end{figure}

To test our heuristic pruning algorithm \ref{algorithm_def}, we have first compared the global minima provided by the 1-SA and 1-PA algorithms against the brute search solution $T_\mathrm{err}^\mathrm{opt}$ of the original index tracking problem defined in \ref{index_tracking_problem}. Figures\ \ref{fig2}a-b compare the value of cost functions of each algorithm with the value of the true optimal solution, for all ratios of tracker size $d$ vs universe size $N$ and multiple dates of our dataset. Points located on the graph diagonal represent scenarios where the heuristic algorithm is able to capture the true optimal solution of the corresponding index tracking problem. The concentration of points along the diagonal is an evidence of the quality of the 1-PA method, which offers a much better heuristic than the 1-SA. We quantify the difference using the Pearson's correlation coefficient, which is $0.92$ and $0.68$ for 1-PA and 1-SA, respectively. Another way to compare both methods is to study the probability that an algorithm finds a portfolio with relative error with respect to the true optimal one $\Delta=(T_\mathrm{err}-T_\mathrm{err}^\mathrm{opt})/T_\mathrm{err}^\mathrm{opt}.$ As shown in figure\ \ref{fig2}c, 1-PA achieves an error $\Delta\leq 20\%$ in $62.5\%$ of the runs, while 1-SA only does it with $17.7\%$ probability. Our analysis indicates that the heuristic pruning is able to incorporate the optimal or close to optimal portfolio for any given size $d.$ 

\begin{figure}[t]
  \centering
  \includegraphics[width=0.7\textwidth]{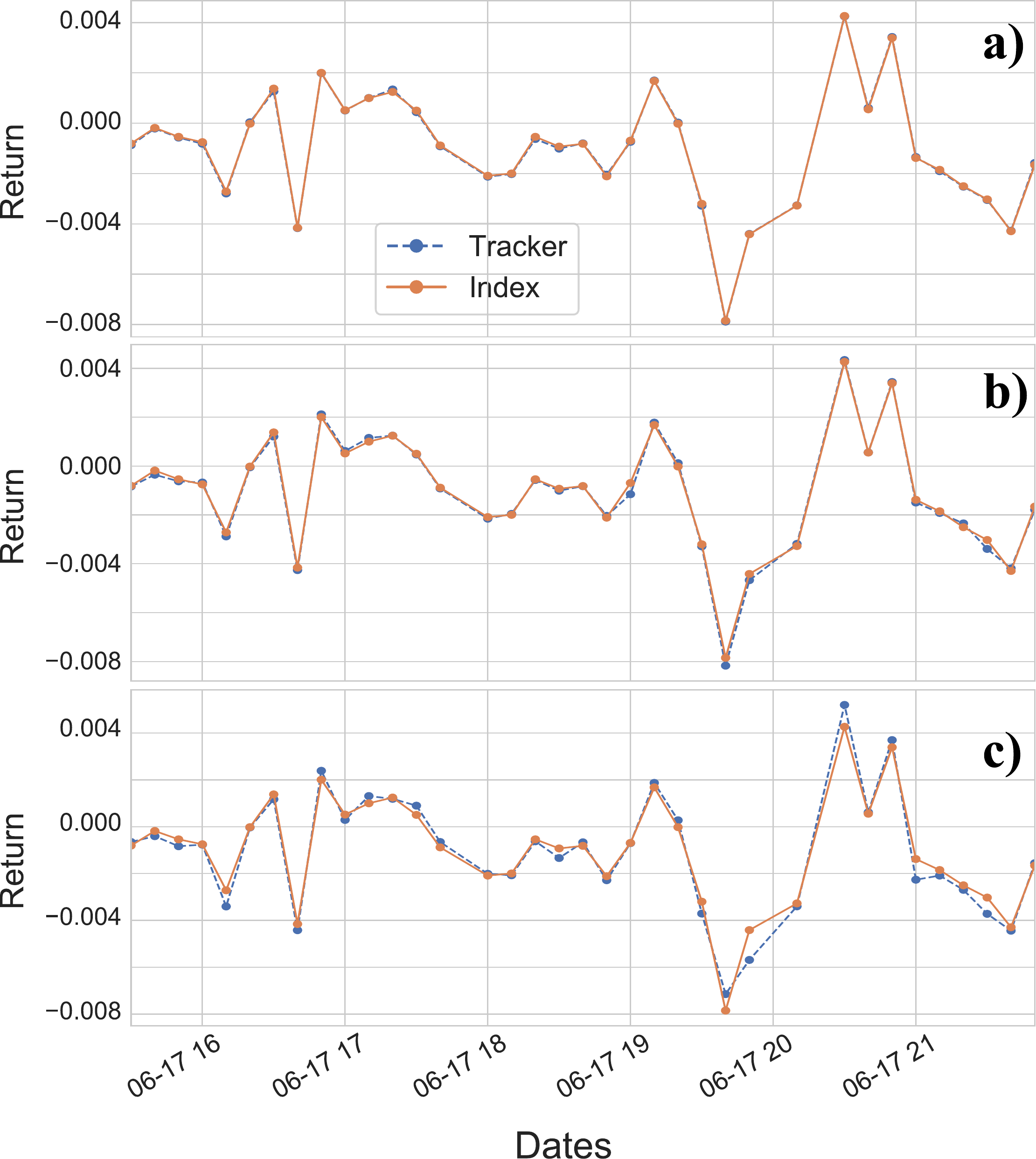}
  \caption{Example of 3-step pruning algorithm with simulated annealing. We have used default parameters from the dimod library \cite{dimod}. Calculations are carried out using $N = 15$ assets taken from the PHLX Oil Service Sector (OSX) index (see Subsection \ref{index_tracking_problem} for details). 
  Benchmark in solid orange line, tracker in dotted blue line. a) First step, reduction from 15 assets to 11. b) Second step, reduction from 11 assets to 7. c) Third step, reduction from 7 assets to 5. Final relative error $\Delta=(T_\mathrm{err}-T_\mathrm{err}^\mathrm{opt})/T_\mathrm{err}^\mathrm{opt}=1.3\%.$}
  \label{fig4}
\end{figure}

We have studied also the performance of the k-PA method, in combination with classical simulated annealing, a well-known classical stochastic algorithm. As publicly available reference, we used the dimod library\ \cite{dimod} for QUBO problems, without fine tuning, but solving the problem with the hard-constraint penalty described in section \ref{hard_constraint}. For a fair comparison, the experiment is designed to arrive to the same portfolio size ($d=5$) with the same number of total overall repetitions (120) through 1-PA $(r_0=120,\alpha=0),$ 2-PA  $(r_0=120,\alpha=4),$ and 3-PA $(r_0=20,\alpha=1)$ respectively.

\begin{figure}[t]
\centering
  \includegraphics[width=0.8\textwidth]{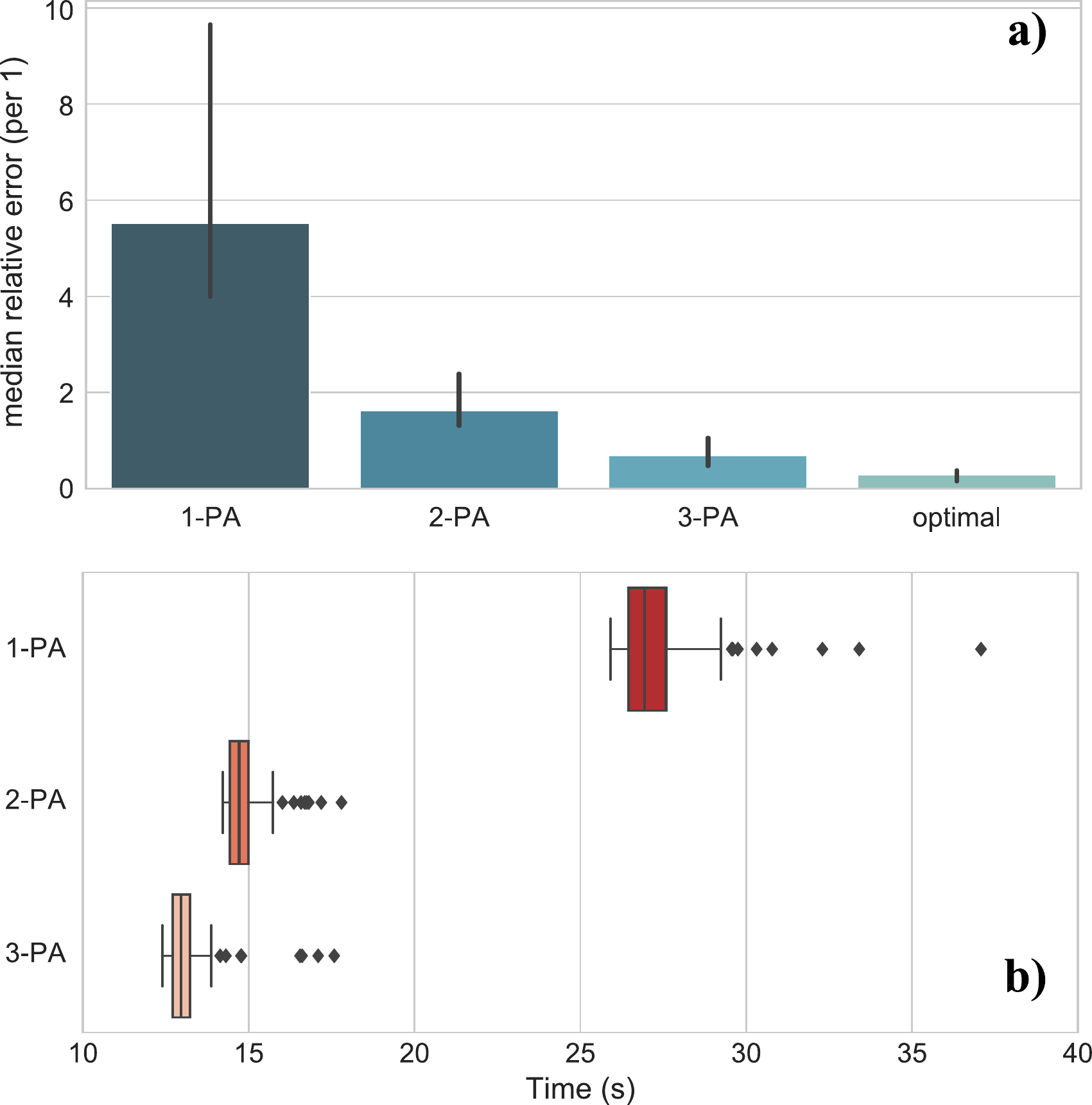}
  \caption{Comparison of multi-step pruning algorithm with simulated annealing for random dates and sizes of the index tracking problem. 
  Calculations are carried out using $N = 15$ assets taken from the PHLX Oil Service Sector (OSX) index (see Subsection \ref{index_tracking_problem} for details)
  a) Bar plot of the median of relative errors $\Delta=\frac{T_\mathrm{err}-T^\mathrm{opt}_\mathrm{err}}{T^{opt}_\mathrm{err}}$ for 1,2,3-step pruning algorithm.
  b) Box plot of the time consumed in each trial. Total number of repetitions is the same for each algorithm 
  (120).}
  \label{fig5}
\end{figure}

The series of figures\ \ref{fig4} illustrates an example of the 3-PA sequence for a random day (2016-06-17), with a final relative error of $\Delta=1.3\%.$ 
The bar plot in figure\ \ref{fig5}a displays a comparison of the median of the relative error $\Delta$ for the multi-step pruning algorithm.
We observe that the error $\Delta$ decreases as we increase the number of steps, $k$. 
This effect can be explained by the fact that a larger $k$ leads to a smaller value $s$ of neglected variables at each step, thus diminishing the error caused by the conversion from continuously valued weights, $w_j$, to binary variables, $x_j$.
The boxplot\ \ref{fig5}b compares the time spent in each algorithm. 
We observe the somehow counter-intuitive effect that a pruning algorithm with more steps takes a longer time to be executed. 
However, note that a larger value of $k$ in $k-$PA
algorithms involves that a lower number of variables, $s$, has to be eliminated at each step. Thus each individual step is defined over a smaller configuration space, which explains the overall speedup effect.
To summarize, our numerical experiment indicates that the k-PA is not only faster as we increase the number of steps, but also more accurate. 

The resulting algorithm offers an alternative appealing heuristic with respect to others local-search algorithms studied in the literature, like the threshold accepting method \cite{gilli2002,gilli2009}, because the combinatorial optimization stage can be tackled by means of several global-search strategies, including quantum optimization heuristics like the ones described in section \ref{variational_states}.

\subsection{Optimization with hard constraints}

\begin{figure}[t]
  \centering
  \includegraphics[width=0.86\textwidth]{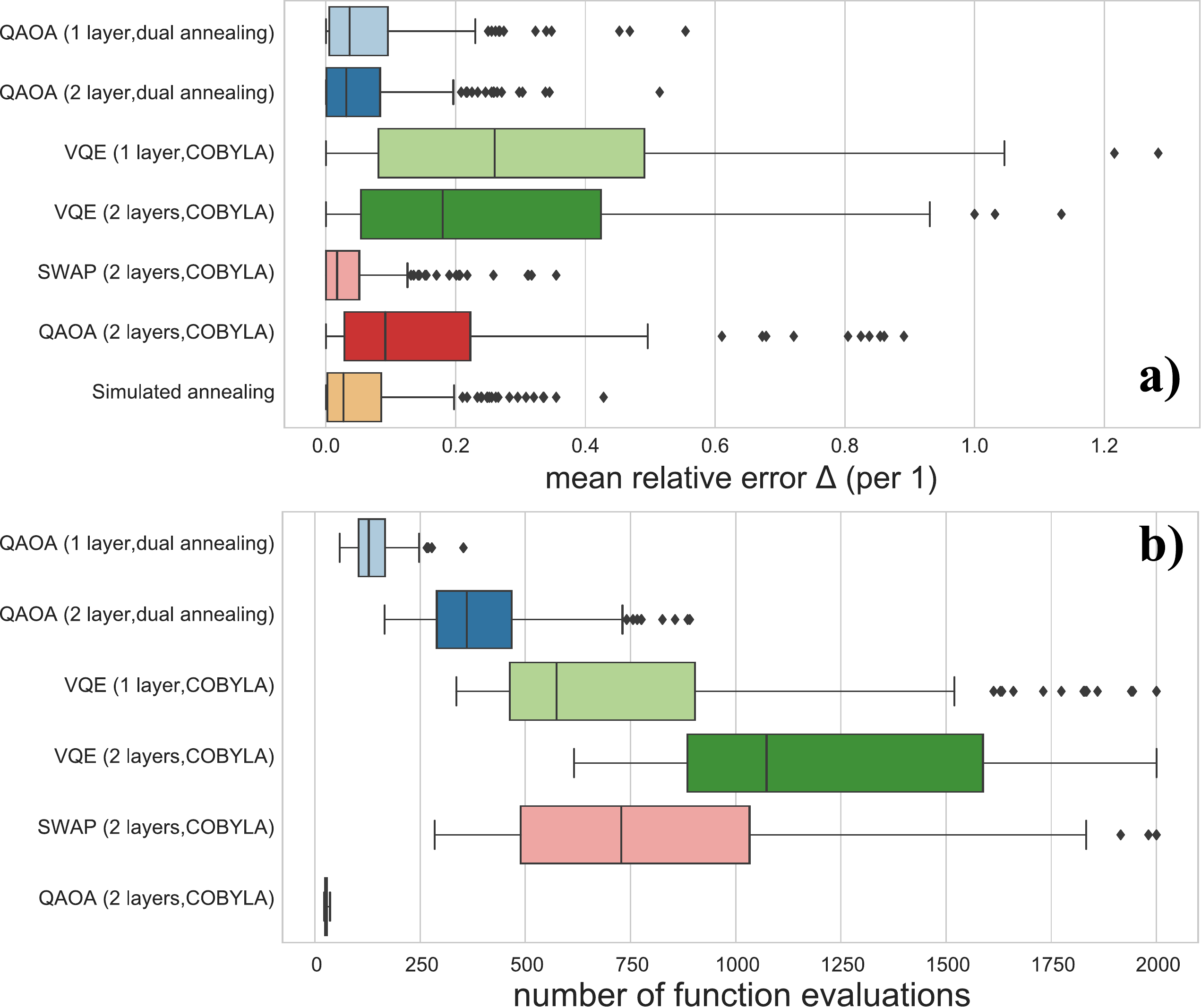}
  \caption{
  Comparison of QAOA, VQE and SWAP ans\"{a}tze, sampling from random dates, random seeds and ratios $d/N$. 
  In our numerical experiments, we select the best solution out of a simulation of measurements $N_{\mathrm{meas}} = 100$.
  Calculations are carried out using $N = 15$ assets taken from the PHLX Oil Service Sector (OSX) index (see Subsection \ref{index_tracking_problem} for details)
  a) Box plot of relative errors 
  $\Delta =
  \frac{P_\mathrm{err}-P^\mathrm{opt}_\mathrm{err}}{P^{opt}_\mathrm{err}}$
  , dimod's simulated annealing with parameter $reads=100$ for a fair comparison 
  with quantum algorithms, where experiments are repeated by 
  $N_\mathrm{meas} = 100$. b) Box plot of number of function evaluations.}
  \label{fig7}
\end{figure}

In solving the combinatorial optimization \eqref{QUBO}, we have compared numerically the performance of  VQE (COBYLA), QAOA (COBYLA), SWAP (COBYLA) and QAOA (dual annealing) using the following procedure. We selected a set of random dates from the whole dataset. For each date we ran each algorithm with a random initial point for every portfolio size (in the case of QAOA the initial point lies within the region previously indicated). 
Once the classical optimization converged, we sampled the variational wavefunction, simulating the outcomes of 100 measurements. 
Out of the resulting binary numbers, we kept the configuration $\mathbf{x}$ with the lowest energy that satisfies the cardinality constraint $\Vert{\mathbf{x}}\Vert_0=d.$ 
As a classical reference, we also calculated the best result among 100 repetitions of simulated annealing using the dimod library.

\begin{figure}[t]
  \centering
  \includegraphics[width=0.7\textwidth]{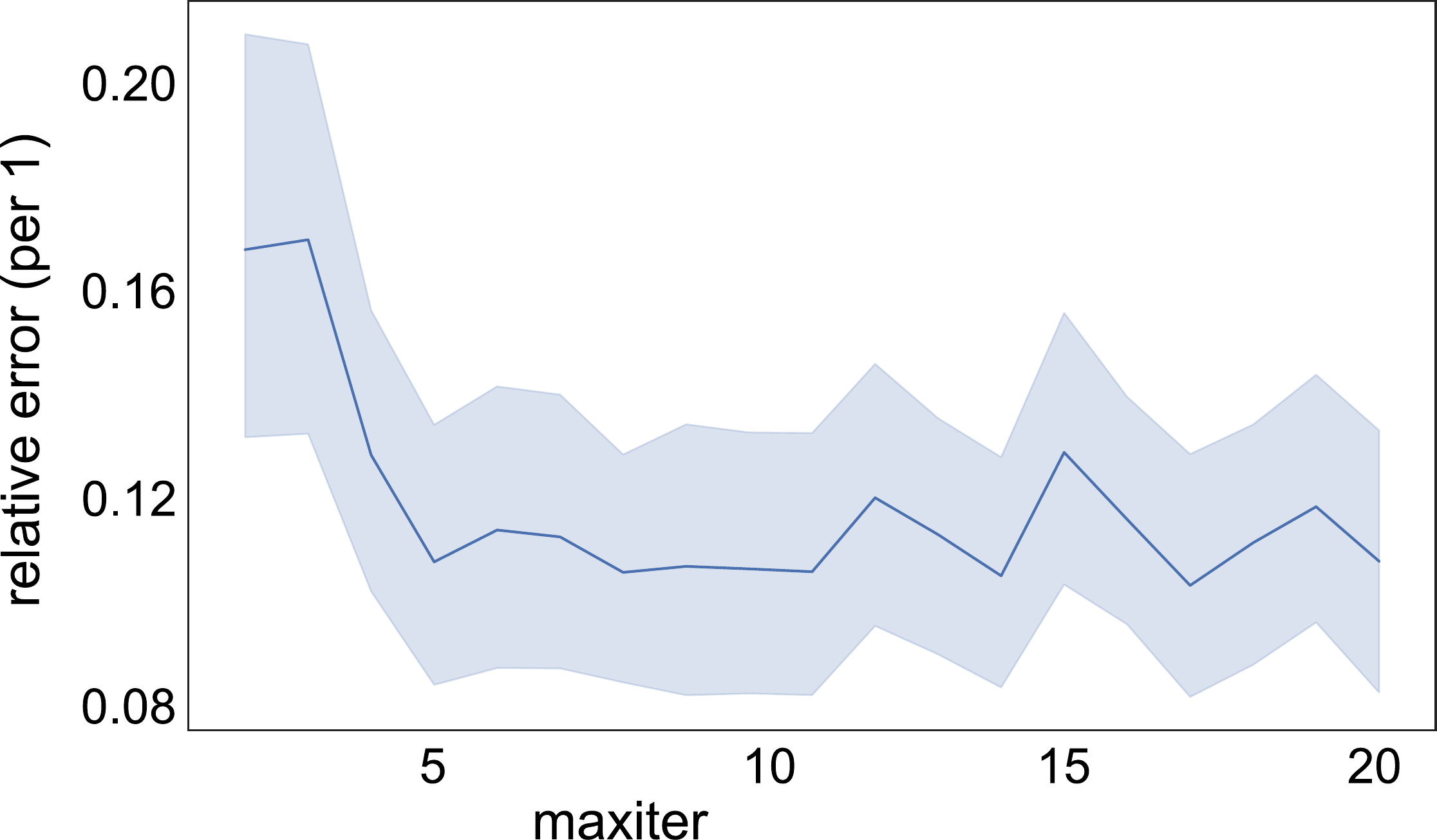}
  \caption{Average relative error $\Delta$ achieved by QAOA with one layer and dual annealing (Scipy) as optimizer. Target portfolio $d=5.$}
  \label{fig6}
\end{figure}

We quantify the performance of the stochastic methods, quantum and classical, using two metrics. Figure\ \ref{fig7}a compares the relative error obtained with different circuits and optimization methods, taking as basis the error of the optimal solution obtained through diagonalization,
$\Delta=\frac{P_\mathrm{err}-P^\mathrm{opt}_\mathrm{err}}{P^{opt}_\mathrm{err}}.$ Figure\ \ref{fig7}b focuses on the number of evaluations for each quantum ansatz and optimization method. In all cases, we explore portfolio sizes in the interval $d = [5,10],$ representing between $1/3$ and $2/3$ of the universe of equities with $N = 15$ (reasonable for the multi-step pruning algorithm). 
Finally, to account for statistical fluctuations, we apply the non-parametric 
Wilcoxon rank-sum test, which tests the null hypothesis that two sets of measurements are drawn from the same distribution---the alternative hypothesis is that values in one sample are more likely to be larger than the values in the other one. Let us point out that the dual annealing method depends on a  $maxiter$
(maximum number of iterations) parameter. 
To attempt a fair comparison with COBYLA, we numerically found a value from which the quality of results plateaus. 
In view of figure \ref{fig6}, that shows the evolution of the relative error for QAOA with one layer and $d = 5$, we conclude
that the method rapidly plateaus, being $maxiter=10$ a reasonable choice.

\begin{figure}[t]
  \centering
  \includegraphics[width=\textwidth]{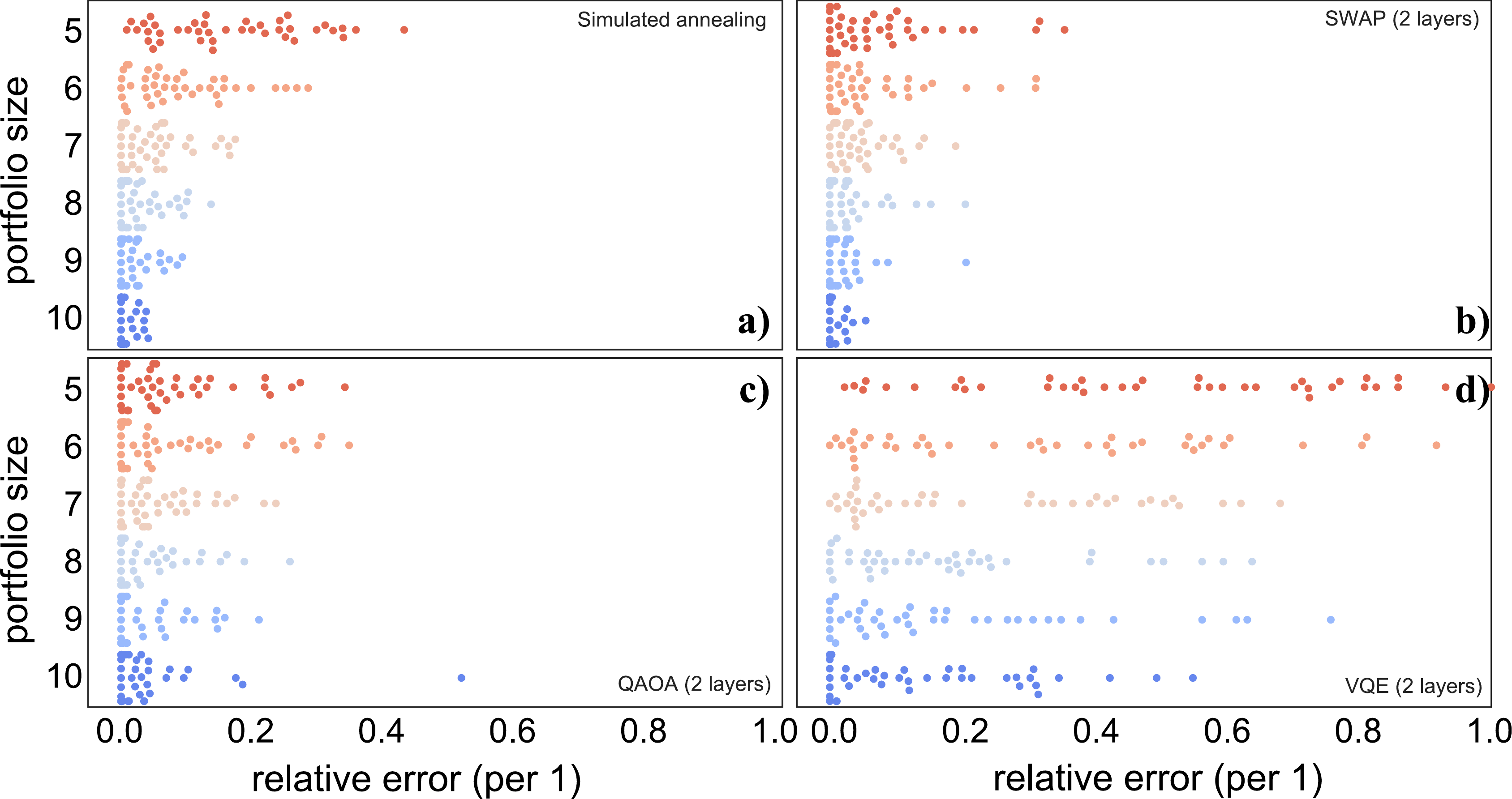}
  \caption{Swarm plot of relative errors $\Delta=\frac{P_\mathrm{err}-P^\mathrm{opt}_\mathrm{err}}{P^{opt}_\mathrm{err}}$ for algorithms that seek a reduction from $N = 15$ to portfolio sizes in the interval $d \in [5,10]$. 
  Calculations are carried out using $N = 15$ assets taken from the PHLX Oil Service Sector (OSX) index
  (see Subsection \ref{index_tracking_problem} for details). 
  We have used: (a) a classical simulated annealing method or a quantum variational method with two layers of (b) SWAP (2 layer, COBYLA), (c) QAOA (2 layer,dual annealing) or (d) VQE (2 layers, COBYLA).
  }
  \label{fig8}
\end{figure}

We can offer some conclusions from the results 
shown in figures \ref{fig7} and \ref{fig6}.
First, the SWAP ansatz offers the lowest average relative error ($\braket{\Delta}=4\%$), followed closely by QAOA with dual annealing ($6\%$). This is confirmed by the rank-sum test with a p-value $p=0.0072$ for a significance level $\alpha=0.05.$ VQE, in comparison, offers poor performance ($31\%$). An additional VQE layer improves the quality, but requires twice the number of function evaluations. We can see that COBYLA quickly gets trapped in local minima for QAOA, leading to worse relative error ($16\%$). 
Finally, an advantage of QAOA over SWAP stems from the fact that it requires less function evaluations to achieve similar relative errors. To summarize, our main conclusion is that, on average, the SWAP ansatz leads to the smallest errors, however, QAOA with dual annealing requires much fewer function evaluations, while still showing reasonably small errors.

We can also look into how the quality of results vary as we increase the ratio $d/N.$ Figure\ \ref{fig8} displays swarm plots for each method whereby we can graphically compare the dispersion of results, their average and the probability of finding the global optimum.
In the swarm plots, the accuracy of the optimization methods 
can be visualized as the concentration of solutions exactly on $\Delta = 0$, which indicates optimal solutions.
All methods are consistent in showing a betterment of all these metrics as the portfolio grows. Interestingly, the probability of finding the optimal portfolio for SWAP and QAOA is better than simulated annealing for the hardest instance ($d=5$); the rank-sum confirms this shift in the distribution with p-values $p=8.3\times 10^{-5}$ and $p=2.7\times 10^{-3}$ respectively. 

In comparison, the test does not find statistical difference for portfolios with a number of assets $d > 7$.

\subsection{Optimization with soft constraints}

\begin{figure}[t]
  \centering
  \includegraphics[width=0.55\textwidth]{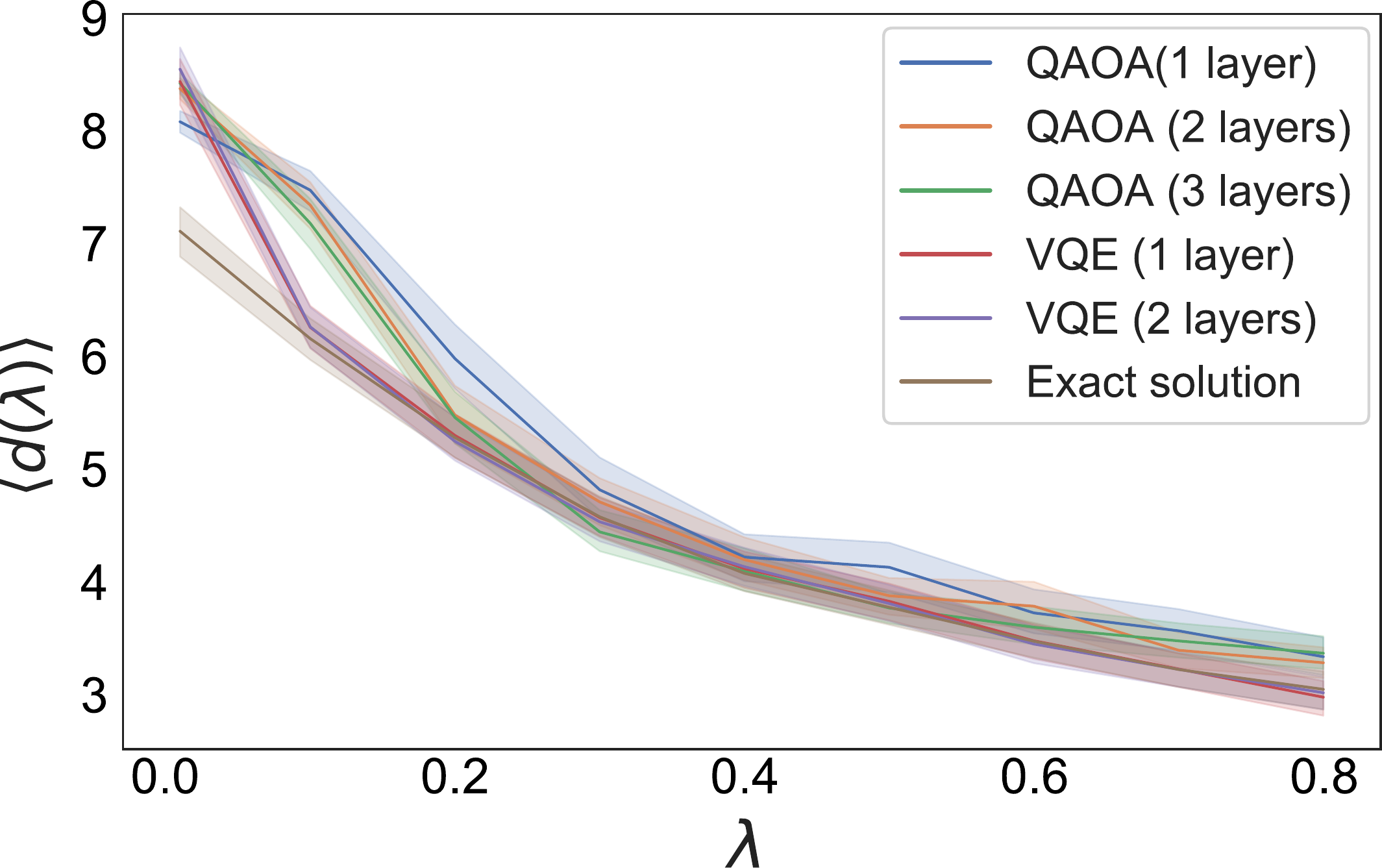}
  \caption{Average portfolio size $\braket{d}$ vs. regularization parameter $\lambda.$ For each value of the regularization parameter, we solve the optimization problem using different ansätze and optimizers. 
  Calculations are carried out using $N = 15$ assets taken from the PHLX Oil Service Sector (OSX) index 
  (see Subsection \ref{index_tracking_problem} for details).
  We compute the average portfolio size by sampling $N_{\rm meas} = $ 100 times the variational wavefunction and using a random choice of dates of the benchmark index.}
  \label{fig9}
\end{figure}

In Sect.\ \ref{soft_constraints} we introduced the idea of a Lagrange multiplier or chemical potential $\lambda$ to enforce a desired value of the soft constraint $d(\lambda).$ This assumed the monoticity of the function $\langle d(\lambda)\rangle$ and the existence of an optimal $\lambda^*$ satisfying $\langle d(\lambda^*)\rangle\approx d_{target}$ for each constraint---in our case, the portfolio size. Therefore, to generate portfolios of a target size $d,$ we must tune\footnote{Remarkably, the values seem to be independent of the ansatz, perhaps suggesting that we could even train a supervised machine learning model that outputs a suitable $\lambda$ given a certain matrix $Q$ so as to automatize the parameter tuning.} $\lambda$ so that the average portfolio produced by our quantum state coincides with $d$ as much as possible. Fig.\ \ref{fig9} indicates that such tuning only needs to happen at the beginning of a series of optimizations over different time windows, because we have found that indices are stable enough to develop a monotonic relation between the regularization parameter and the average portfolio size. For a given $\lambda,$ the average size remains fairly stable for different random initial points and different dates. 

Small deviations from the optimal $\lambda^*$ will not cause a failure to recover the right portfolios---they may affect the quality of the results and the likelihood of finding low energy configurations with the right size $d$---, and one might also design search strategies by which the value of $\lambda$ may be dynamically adjusted.

\begin{figure}[t]
  \centering
  \includegraphics[width=\textwidth]{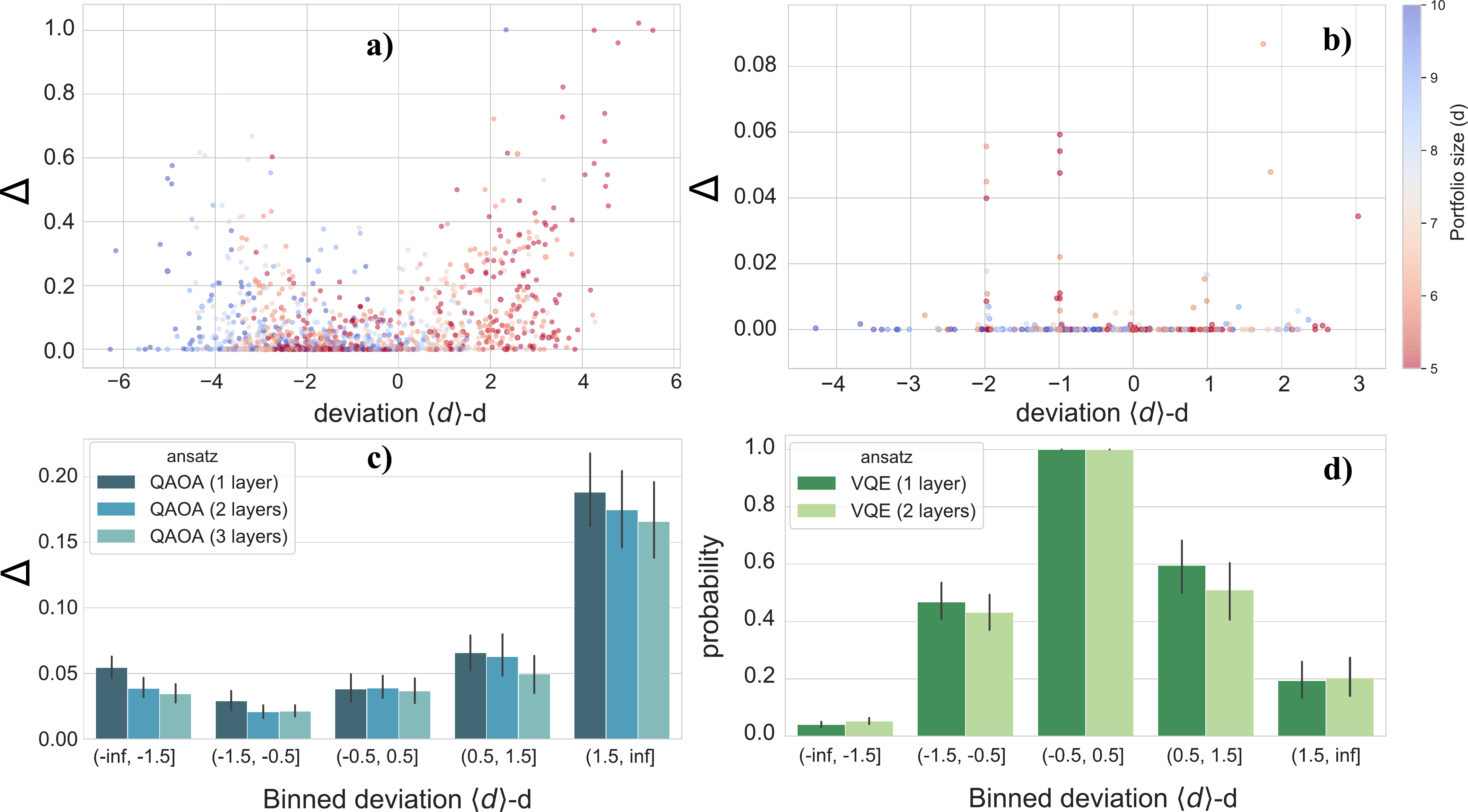}
  \caption{Soft constraint variational optimization performance with QAOA and VQE. Relative error vs. deviation in portfolio size $\langle d\rangle -d$ for (a) QAOA or (b) VQE, both with one layer, solved using and COBYLA. (c) Relative error ($\Delta=\frac{P_\mathrm{err}-P^\mathrm{opt}_\mathrm{err}}{P^{opt}_\mathrm{err}}$) vs. binned deviation in portfolio size, for QAOA with 1, 2 and 3 entangling layers. (d) Probability of success at sampling a portfolio of size $d,$ for VQE with 1 and 2 entangling layers. Calculations are carried out using $N = 15$ assets taken from the PHLX Oil Service Sector (OSX) index 
  (see Subsection \ref{index_tracking_problem} for details). $N_{\rm meas} = 100$.}
  \label{fig10and11}
\end{figure}

To illustrate this effect, in figure\ \ref{fig10and11}a we present a scatter plot of relative errors achieved by QAOA with 1 layer and COBYLA as a function of the difference between the mean $\langle d\rangle$ and the desired size $d$ (the baseline here is the hard-constraint exact solution (calculated by exact brute-force search) for a fair comparison with previous section); a binned version is shown in plot\ \ref{fig10and11}c, also comparing the impact of additional layers. We employed the same numerical procedure described earlier, but in the last step we select the portfolio with the lowest energy satisfying the cardinality constraint from the 100 measurements (if such configuration exists). Effectively, the relative error progressively worsens as we separate from $\lambda^*,$ yet we can expect good results within two units around $\lambda^*.$ The asymmetry stems from the fact that it is more likely to find smaller $d's$ on the right hand of the optimal point, which statistically yields greater dispersion as seen in the previous section. We can also see that introducing layers slightly enhance the results; a rank-sum test identifies a conclusive difference between the first and second layer distribution with p-value $p=1.5\times 10^{-4},$ and a somewhat less conclusive $p=0.024$ between the second and third layer. 

A strikingly contrasting scenario shows up when repeating the same analysis for VQE, as the scatter plot in Figure\ \ref{fig10and11}b reveals: sampling the global optimum is very likely to happen as long as VQE succeeds in returning a configuration satisfying the cardinality constraint. Deterioration is precisely noticed in the probability of reaching such feasible portfolio, rapidly decaying as we deviate from the optimal $\lambda^*$ as plotted in figure\ \ref{fig10and11}d (unlike QAOA, which remains fairly stable for greater deviations). In essence, VQE almost guarantees to pinpoint the global optimum provided that we are close enough to the optimal regularization parameter $\lambda\simeq\lambda^*.$ Introducing an additional layer does not seem to provide any advantage (the rank-sum tests throws an inconclusive $p=0.54$). The main takeaway from our analysis in Fig.\ \ref{fig10and11} is, thus, that QAOA seems more robust to deviations from the optimal parameter, whereas VQE requires, on average, a better tuning of this parameter. 

\begin{figure}[t]
  \centering
  \includegraphics[width=0.7\textwidth]{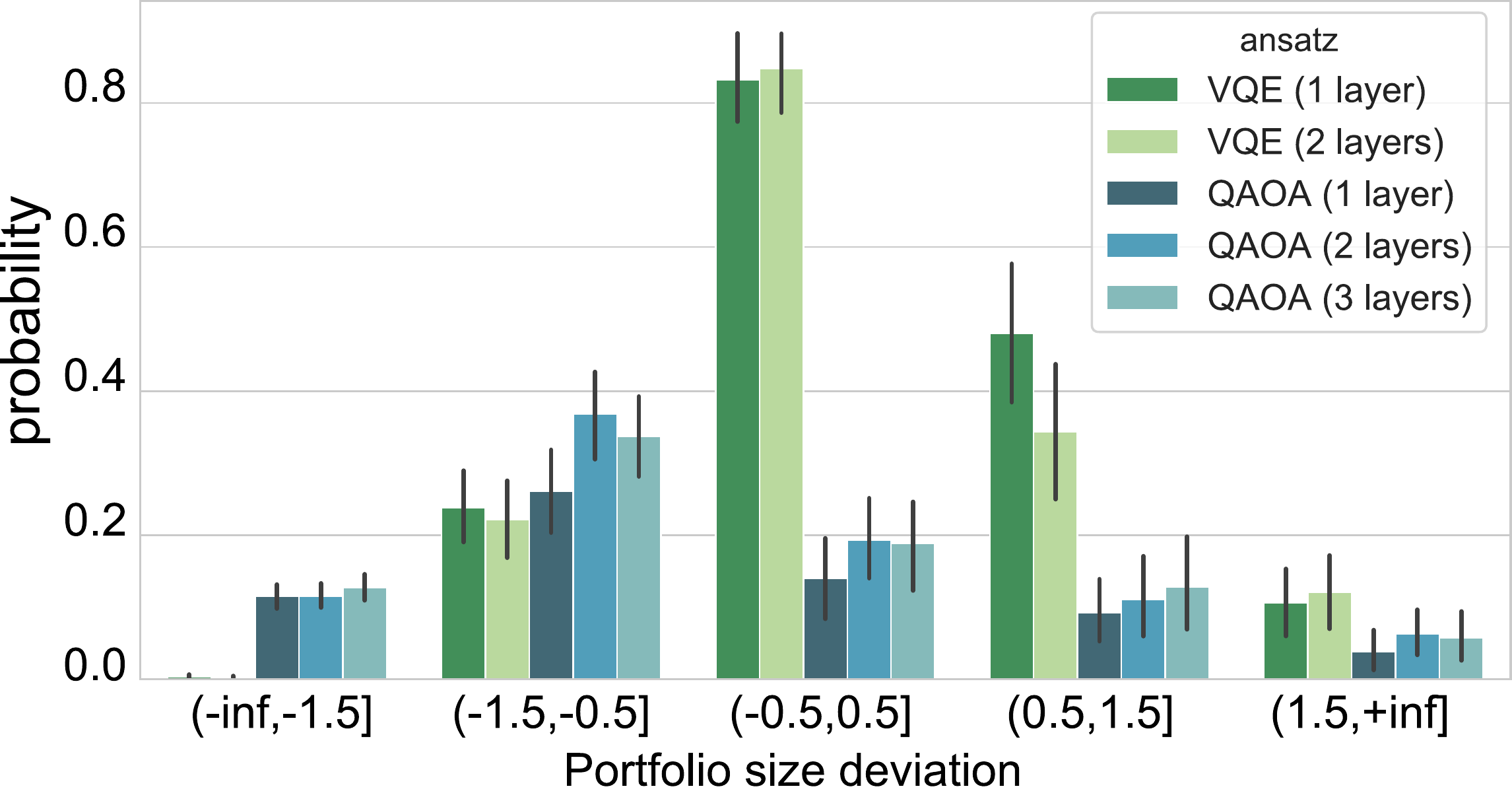}
  \caption{Probability of sampling the global optimum as a function of the binned deviation  $\langle d\rangle -d,$ comparing various ans\"atze and number of entangling layers. Calculations are carried out using $N = 15$ assets taken from the PHLX Oil Service Sector (OSX) index 
  (see Subsection \ref{index_tracking_problem} for details), and $N_{\rm meas} = 100$.}
  \label{fig12}
\end{figure}

\begin{figure}[t]
  \centering
  \includegraphics[width=0.8\textwidth]{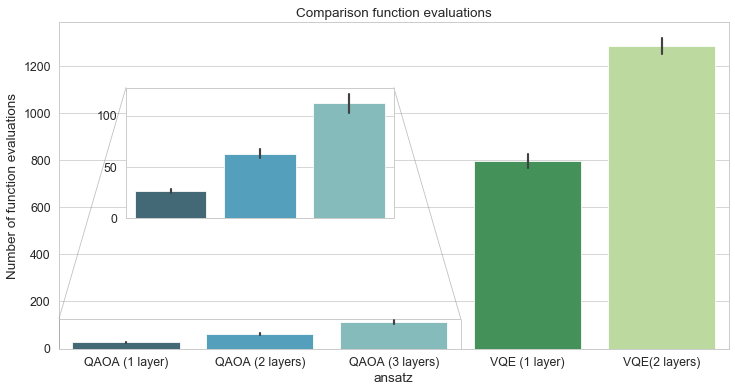}
  \caption{Bar plot of average number of function evaluations, comparison between QAOA and VQE with $|\langle d\rangle -d|<1$.
  Calculations are carried out using $N = 15$ assets taken from the PHLX Oil Service Sector (OSX) index 
  (see Subsection \ref{index_tracking_problem} for details), and $N_{\rm meas} = 100$.}
  \label{fig13}
\end{figure}

A metric that somehow summarizes our previous discoveries is the probability of sampling the global optimum, compared in the bar plot in figure \ \ref{fig12} for both VQE and QAOA. VQE clearly outstrips QAOA in this regard, but when comparing the number of function evaluations (bar plot in figure\ \ref{fig13}) we note that it requires around 8 times more calls to the quantum processor. In the light of these results, we conclude that the soft-constraint QUBO encoding improves enormously the energy landscape with respect to the hard-constraint version, leading to better metrics. Provided we are close enough to the optimal regularization parameter $\lambda^*,$ QAOA improves performance in both relative error and calls even when using COBYLA, where previously we had to rely on a global optimizer. VQE offers high probability of reaching the global optimum but is less robust to large deviations from  $\lambda^*$ than QAOA, and it also demands many more function evaluations.

We have analyzed results for a single pruning step, but the soft-constraint formulation can also be merged into the multi-step algorithm. Since it is not necessary to keep an exquisite control over the portfolios sizes at intermediate steps, a pruning schedule may include a schedule also for the regularization parameters $\lambda_0\geq \lambda_1\geq \dots\geq \lambda_k.$ We would start from a high enough value $\lambda_0$ to explore larger portfolios, and carefully select the regularizers close to the end, so that $\lambda_k\to\lambda^*.$ Another idea is to apply different ansatz and constraint formulations at different stages of the pruning. One could use a VQE-like ansatz with a soft encoding for the intermediate steps, where no fine tuning of $\lambda$ is needed, and switch to a low-depth QAOA with hard constraints at the end, to have control over the final portfolio size $d.$

\section{Conclusions} \label{conclusions}
In this work we have introduced an iterative algorithm to solve continuous optimization problems with a cardinality constraint that assimilates their difficulty to combinatorial optimization problems. The algorithm  iteratively discards variables through a constrained QUBO optimization process, which is driven by quantum optimization algorithms such as QAOA.

\begin{figure}[t]
  \centering
  \includegraphics[width=0.8\textwidth]{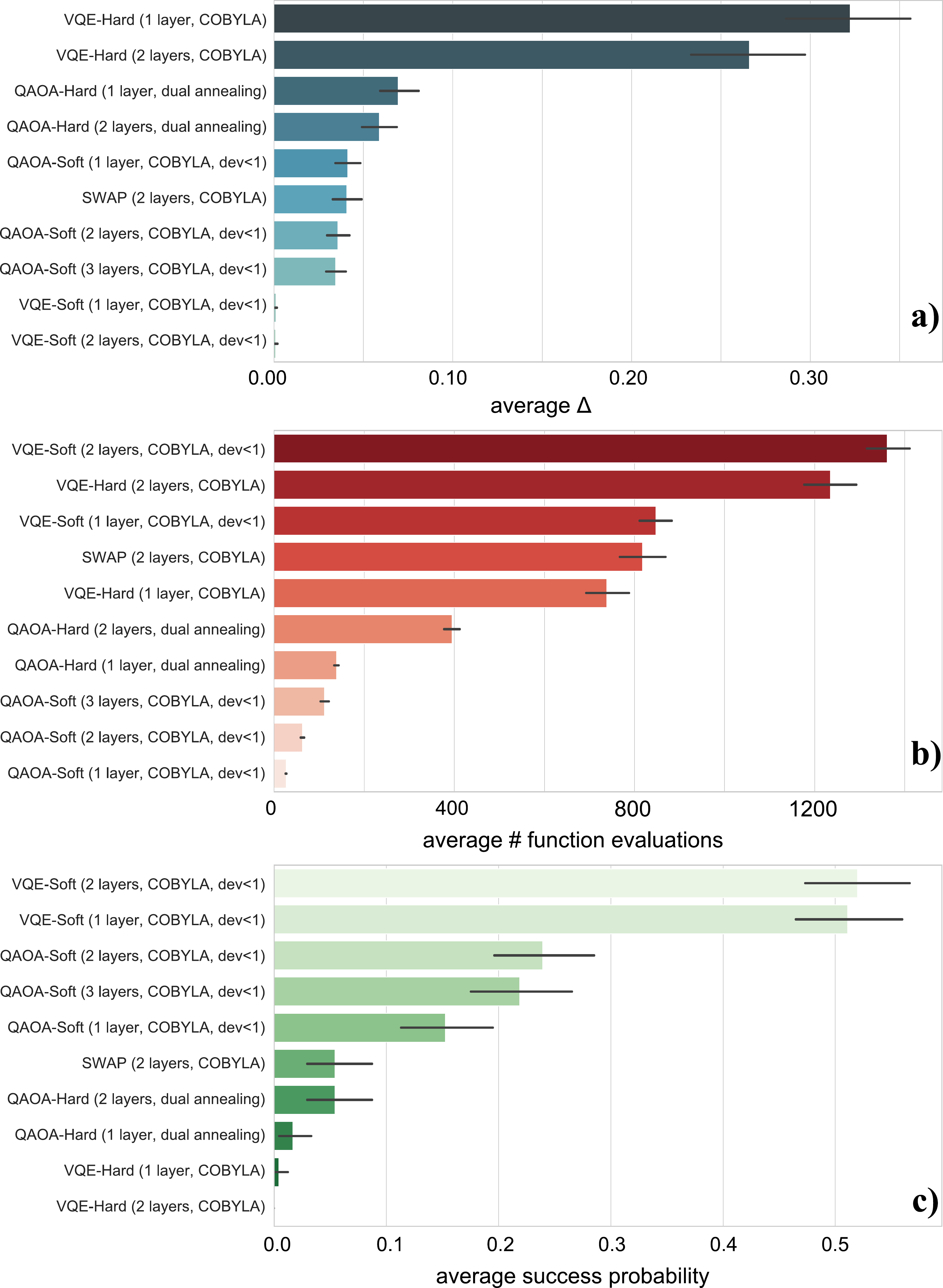}
  \caption{Bar plots representing a ranking of methods based on (a) average of relative errors $\Delta=\frac{P_\mathrm{err}-P^\mathrm{opt}_\mathrm{err}}{P^{opt}_\mathrm{err}}$ (b) average number of function evaluations and (c) probability of sampling the global optimum (where dev$<1$ stands for $|\langle d\rangle -d|<1$ as for soft constraint methods). Calculations are carried out using $N = 15$ assets taken from the PHLX Oil Service Sector (OSX) index 
  (see Subsection \ref{index_tracking_problem} for details), and $N_{\rm meas} = 100$.}
  \label{fig14}
\end{figure}

Through numerical experiments with the financial index problem as a workbench, we found that the performance of the pruning algorithm depends on the combination of three factors: the quantum ansatz, the classical optimizer, and the mathematical encoding. Figure \ref{fig14} displays a ranking of all the methods studied regarding averages obtained for three key metrics: the relative error with respect to the exact solution (calculated by exact brute force search), the number of function evaluations and the probability of sampling the global optimum.

As we can see from figure \ref{fig14}, the mathematical encoding is the most prominent actor determining the performance for both error and sampling. Introducing penalty terms in a QUBO formulation to account for cardinality constraints drastically increases the hardness of the problem. Soft constraints or quantum circuits designed to search in appropriate sub-spaces become practical alternatives. Local rotations in the VQE ansatz greatly decreases the speed of the algorithm, but it may also lead to high probability of sampling the optimum. QAOA seems well-balanced between speed and accuracy.

Our results hopefully extend the potential applications of NISQ devices in the near future. Beyond the finance industry, the tools developed in this work may impact other fields. For instance, it can be applied for selecting variables in a multi-variate linear regression \cite{miller2002} (a common problem in machine learning and econometrics) with a different heuristic with respect to local-search algorithms like threshold accepting \cite{zilinskas2003,gilli2019}, as an alternative to improve ridge regularizations \cite{lin10,johnson15}. Other applications in machine learning include the sparse Principal Component Analysis \cite{d2005direct}; in ensemble machine learning, it also serves as an extension of the QBoost algorithm\ \cite{neven2008} for selecting $K$ weak learners out of a collection of $N$ predictors. Our results could also find applications in compressed sensing/sampling for signal processing \cite{candes2008}, or the capacitated facility location problem for supply chain \cite{velasquez2019}.

\section*{Acknowledgments}
D. Porras and J.J. Garc\'\i a Ripoll acknowledge funding from Spanish project PGC2018-094792-B-I00  (MCIU/AEI/FEDER, UE),  CSIC Research Platform
PTI-001, and CAM/FEDER Project No. S2018/TCS4342 (QUITEMAD-CM). We thank Ecol\'astico Sanchez for fruitful discussions about the index tracking problem.

\section*{References}
\bibliographystyle{unsrt}
\bibliography{biblio}

\end{document}